\RequirePackage{fix-cm}
\documentclass[twocolumn]{svjour3_arxiv}
\smartqed
\usepackage{graphicx}
\usepackage{amsfonts}
\usepackage{amsmath}
\usepackage{natbib}
\usepackage[usenames,dvipsnames]{color}
\usepackage{soul}

\newcommand{\be}{\begin{equation}}
\newcommand{\ee}{\end{equation}}

\newcommand{\gf}{\mbox{{$\boldsymbol f$}}}
\newcommand{\gn}{\mbox{{$\boldsymbol n$}}}

\newcommand{\gtau}{\mbox{$\boldsymbol\tau$}}
\newcommand{\gu}{\mbox{{$\boldsymbol u$}}}
\newcommand{\gxi}{\mbox{$\boldsymbol\xi$}}
\newcommand{\gzero}{\mbox{$\boldsymbol 0$}}

\journalname{arXiv.org}
\begin{document}

\title{
A Note on Optimal Design of Multiphase Elastic Structures
\thanks{}
}

\author{Nathan Briggs \and Andrej Cherkaev \and Grzegorz Dzier\.z{}anowski}

\institute{
N. Briggs \at
Department of Mathematics, University of Utah\\
155S 1400E, Salt Lake City, UT 84112, USA\\
\email{nathancbriggs@gmail.com}
\and
A. Cherkaev \at
Department of Mathematics, University of Utah\\
155S 1400E, Salt Lake City, UT 84112, USA\\
Tel.: +1-801-581-6822/6851\\
Fax: +1-801-581-4181\\
\email{cherk@math.utah.edu}
\and
G. Dzier\.z{}anowski \at
Faculty of Civil Engineering, Warsaw University of Technology\\
al. Armii Ludowej 16, 00-637 Warsaw, Poland\\ 
Tel.: +48-22-2346516\\
Fax: +48-22-2345095\\
\email{gd@il.pw.edu.pl}
}

\date{Received: date / Accepted: date}

\maketitle

\begin{abstract}
The paper describes the first exact results in optimal design of three-phase elastic structures. Two isotropic materials, the ``strong'' and the ``weak'' one, are laid out with void in a given two-dimensional domain so that the compliance plus weight of a structure is minimized. As in the classical two-phase problem, the optimal layout of three phases is also determined on two levels: macro- and microscopic. On the macrolevel, the design domain is divided into several subdomains. Some are filled with pure phases, and others with their mixtures (composites). The main aim of the paper is to discuss the non-uniqueness of the optimal macroscopic multiphase distribution. This phenomenon does not occur in the two-phase problem, and in the three-phase design it arises only when the moduli of material isotropy of ``strong'' and ``weak'' phases are in certain relation.

\end{abstract}

\section{Introduction}
\label{sec:intro}
Multiphase elastic structures designed for minimal compliance are made from composites with microgeometries of maximal stiffness. In the classical pro\-blem of \emph{two-phase optimal design} analyzed in the framework of two-dimensional elasticity, rank-1 or rank-2 laminates are the examples of optimal mixtures. In the latter, the ``strong'' material envelopes the ``weak'' one; concentration of the ``strong'' phase in a rank-2 laminate grows with the intensity of the average stress (also referred to as homogenized, or effective stress); and the anisotropy of a microstructure follows the anisotropy of the effective stress tensor, see \citep{Lur82, Gib87, Ben88, Che00}. Another type of optimal two-phase layout corresponds to the confocal ellipse construction, see \citep{Gra95}. However, it is worth pointing out here that rank-2 laminates are uniquely optimal in 2D elasticity when the sign of the determinant of the homogenized stress tensor is negative, see \citep{All99}. The same phenomena are observed in the limiting case of \emph{topology optimization}, when the ``weak'' material degenerates to void, see \citep{Vig94, All02, Ben03}. 

Optimal multiphase composites are much less investigated; notice the pioneering contributions of  \citet{Gib00} and \citet{Sig00}, see also \citep{Alb07, Che11, Che13} for continuation and extensions. 

In the present paper we discuss the first exact generalization of both the two-material and material-void problems. More precisely, our concern is to minimize the compliance of a two-dimensional linearly elastic stru\-ctu\-re made from three phases - two isotropic materials and a void. The focus and main aim of the discussion is to elaborate in detail the case when optimal macroscopic distribution of materials in a multiphase structure is not unique. This feature is special as it does not occur in two-phase problems of optimal design. 

Subsequent considerations regarding the optimality of multiphase microstructures are focused on high-rank orthogonal laminates. However, other choices are also possible; e.g.\ the three-phase wheel assemblages studied in \citep{Che12} are proven to be optimal if the homogenized stress tensor is isotropic. 

Closely related to the problem of optimal structural design are the works on bounding the properties of multimaterial mixtures, see e.g. \citep{Mil90, Nes95}. For extensive exposition of this topic and list of references we refer the reader to the monograph by \citet{Mil02}.

\section{The problem}
\label{sec:problem}
\subsection{Notation}
\label{sec:notation}
For simplicity we assume that ``strong'' and ``weak'' elastic materials are isotropic with Poisson coefficients equal to zero. It follows that bulk and shear moduli of each material are equal. However, the results can be easily generalized to arbitrary well-ordered phases, that is to the case when both the bulk and shear modulus of one phase is greater than the corresponding modulus of the other. 

Let $C_1$ and $C_2$, $C_1< C_2$, denote material compliances (inverses of Young's moduli); the compliance of void is infinite. Define
\be
\label{eq01}
C(x)=\begin{cases}
C_1 &\mbox{if } x\in \Omega_1, \\
C_2 &\mbox{if } x\in \Omega_2,\\
+\infty &\mbox{if } x\in \Omega_3,
\end{cases}
\ee
where  $\Omega_1$, $\Omega_2$ and $\Omega_3=\Omega_{\rm{void}}$ are disjoint subdomains in a bounded domain $ \Omega$ occupied by phase 1 (strong material), phase 2 (weak material) and phase 3 (void) respectively.

Equilibrium conditions and constitutive equations of linearized elasticity are
\be
\label{eq02}
\nabla \cdot \gtau = \gzero\mbox{ in }\Omega,\ \ \gtau \, \gn = \gf  \mbox{ on } \partial \Omega_f,\ \ \gtau=\gtau^{\rm{T}}
\ee
\be
\label{eq03}
2C \gtau = \nabla\gu + \nabla\gu^{\rm{T}},\ \ \gu = \gu_0 \mbox{ on } \partial \Omega_u, 
\ee
where $\gtau$  is a $2 \times 2$ tensor stress field,  $\gu$ is a vector displacement field, $\gf$ denotes traction on the boundary component $\partial \Omega_f$, $\gu_0$ stands for a trace of displacement field on $\partial \Omega_u$ and $\gn$ is a normal to the boundary.

Stress energy accumulated in $i$-th material, $i=1, 2$, is given by 
\be
\label{eq04}
2\,W_i(\gtau) =  C_i\,\mbox{Tr}\,\big(\gtau^2\big),
\ee
and for void (phase 3) it is assumed that 
\be
\label{eq04a}
W_3(\gtau)=\begin{cases}0 &\mbox{if } \gtau=\gzero, \\ +\infty &\mbox{otherwise} . \end{cases}
\ee

\noindent\emph{Remark}:\\
In the sequel we identify a stress tensor $\gtau$ with its eigenvalues and write $\gtau=(\tau_{\rm{I}}$, $\tau_{\rm{II}})$.

\subsection{The optimization problem}
\label{sec:optim}
Consider the optimization problem: among all divisions of $\Omega$ into disjoint subdomains $\Omega_i$, $i=1, 2, 3$, whose areas $|\Omega_i|$ are restricted by $|\Omega_i|\le V_i$, choose the one that minimizes structural compliance. Here $V_i$ denotes the maximal amount of the $i$-th phase that can be used to compose an optimal structure. This problem can be formulated in a variational form
\be
\label{eq05}
I = \inf \bigg\{\int_\Omega F(\gtau)\,dx\ \bigg|\ \gtau\mbox{ as in \eqref{eq02}}\bigg\}
\ee
where
\be
\label{eq06}
F (\gtau)= \begin{cases}
0 &\mbox{if } \gtau=\gzero, \\
\min\big\{\Phi_1(\gtau),\ \Phi_2(\gtau)\big\} &\mbox{otherwise}.
\end{cases}
\ee
with
\be
\label{eq06a} 
\Phi_i(\gtau) = 2\,W_i(\gtau) + \gamma_i,\quad i=1, 2, 
\ee
denoting \emph{the energy well of $i$-th material}. Here, $\gamma_i$ stand for the Lagrange multipliers related to the restrictions on $|\Omega_i|$ set above. They can be understood as ``costs'' of materials; in the sequel we assume $\gamma_2=\alpha\,\gamma_1$ with $\alpha\in (0,1)$ denoting the quotient of material costs.

The integrand $F$ is not a convex function of $\gtau$, hence the variational problem \eqref{eq05} has no classical solution in the sense that an optimal division of $\Omega$ into three disjoint subdomains occupied by pure phases does not exist in general. The original optimization problem requires \emph{relaxation} by allowing arbitrary mi\-cro\-stru\-ctu\-ral mixtures of pure pha\-ses (limits of classical solutions) in optimal design. Effective constitutive properties of thus obtained composite materials are determined by \emph{the homogenization theory}, see e.g.\ \citep{Che00, All02}.

Technically, relaxation results in replacing $F$ with its \emph{quasiconvex envelope} $QF$, see \citep{Dac08}. Due to the local character of homogenization, formulae for $QF(\gtau)$ and the properties of related optimal microstructures can be determined independently at each point of the design domain $\Omega$. The original optimization problem \eqref{eq05} is consequently replaced by
\be
\label{eq06b}
QI = \min\bigg\{\int_\Omega QF(\gtau)\,dx\ \bigg|\ \gtau\mbox{ as in \eqref{eq02}}\bigg\}
\ee
with
\be
\label{eq06c}
\begin{array}{l}
QF(\gtau) = \\
= \min\bigg\{2\,W^\ast(\gtau,m_1,m_2,m_3)+\gamma_1\,m_1+\alpha\gamma_1\,m_2 \\
\bigg|\ 0\le m_i\le 1,\ i=1,2,3,\ m_1+m_2+m_3=1\bigg\}
\end{array}
\ee
where $m_1, m_2, m_3$ respectively denote macroscopic volume fractions of the strong, weak, and void phases making up the composite and $W^\ast$ stands for the stress energy density accumulated in this mixture.  It is worth pointing out that $QF(\gtau)\le F(\gtau)$ for any $\gtau$ satisfying \eqref{eq02}, and $QI = I$.

\section{Study of the quasiconvex envelope}
\label{sec:env}
In the problem considered here, $QF$ is supported by the energy wells $\Phi_1$, $\Phi_2$ corresponding to strong and weak materials and a well of the void energy, see \eqref{eq04a}. Macroscopic volume fractions of phases in the locally optimal composite are still denoted by $m_i$, $i=1, 2, 3$. The optimal values of $m_i$ depend on the quotient of material costs $\alpha$, and eigenvalues of the homogenized stress $\gtau$.

In the following analysis, we fix the cost $\gamma_1$ of the strong material and study the shape of the quasiconvex envelope in dependence on the ratio $\alpha=\gamma_2/\gamma_1$.

Let $\mathsf{R}(\alpha)$ be the set of stresses for which the energy accumulated in a composite plus its ``cost'' has lower value than for any of the pure materials: $\mathsf{R}(\alpha)=\{\gtau : QF(\gtau)<F(\gtau)\}$ for given $\alpha$. If $\gtau\in \mathsf{R}(\alpha)$ then the minimal stress energy is accumulated in a composite material and $m_i<1$, $i=1,2,3$. Conversely, if $\gtau\notin \mathsf{R}(\alpha)$ then $QF(\gtau)=F(\gtau)$ and the energy is minimized on pure $i$-th phase that is $m_i=1$, $i\in\{1,2,3\}$. In this sense we refer to $\mathsf{R}(\alpha)$ as \emph{the composite region} whose topology depends on the quotient of material costs.

Below we briefly discuss the changes in the topology of $\mathsf{R}(\alpha)$ with regard to $\alpha$. In particular, we elaborate on the special case when $\alpha$ takes such a value for which $m_i$ are non-unique for all $\gtau\in\mathsf{R}(\alpha)$, see Sec.\ \ref{sec:critical2} below.

\subsection{Large quotient of material costs}
\label{sec:high2}
Consider the case when the quotient of material costs is greater than a certain threshold, $\alpha>\alpha_A$ (see Sec.\ \ref{sec:critical2} for details), and material 2 is not used in optimal design at all, i.e.
\be
\label{eq07}
m_2=0. 
\ee
As in the topology optimization problem, see \citep{Che00, All02}, the quasiconvex envelope is supported by the energy well $\Phi_1$ and $W_3(\gzero)=0$. For $\gtau = (\tau_{\rm{I}}, \tau_{\rm{II}})$ we thus have $QF = QF_{13}$,
\be
\label{eq08}
\begin{array}{l}
QF_{13}(\gtau) = \\
=\left\{
\begin{array}{ll}
2\,C_1\big[\xi_0\big(|\tau_{\rm{I}}|+ |\tau_{\rm{II}}|\big) - |\tau_{\rm{I}}\,\tau_{\rm{II}}|\big] &\mbox{   if } \gtau\in \mathsf{R}(\alpha),\\[4pt]
C_1 (\tau_{\rm{I}}^2+ \tau_{\rm{II}}^2) +\gamma_1  &\mbox{   otherwise} ,
\end{array}\right.
\end{array}
\ee
with 
\be
\label{eq08a}
\mathsf{R}(\alpha)=\{\gtau : |\tau_{\rm{I}}|+ |\tau_{\rm{II}}| \leq\xi_0\}, \quad\xi_0 = \sqrt{\dfrac{\gamma_1}{C_1}}. 
\ee
For $\gtau\in\mathsf{R}(\alpha)$ optimal composites made from phase 1 (strong material) and phase 3 (void) can take a form of rank-2 orthogonal laminates $L(13,1)$ with macroscopic volume fractions given by
\be
\label{eq08b}
 m_1=\dfrac{|\tau_{\rm{I}}| + |\tau_{\rm{II}}|}{\xi_0},\quad m_3=1-m_1.
\ee
From \eqref{eq07} and \eqref{eq08b} it follows that $m_1$, $m_2$ and $m_3$ are uniquely determined for all $\gtau\in\mathsf{R}(\alpha)$ if $\alpha$ is large enough.

Components of $\gtau^{(1)}$ and $\gtau^{(3)}$ satisfy pointwise relations
\be
\label{eq09}
|\tau_{\rm{I}}^{(1)}| + |\tau_{\rm{II}}^{(1)}| =  \xi_0 \mbox{     in }\Omega_1 ,\quad
\tau_{\rm{I}}^{(3)} = \tau_{\rm{II}}^{(3)} = 0 \mbox{     in }\Omega_3 .
\ee

\subsection{Special quotient of material costs}
\label{sec:critical2}
 
\begin{figure}
\centering
\includegraphics[scale=1]{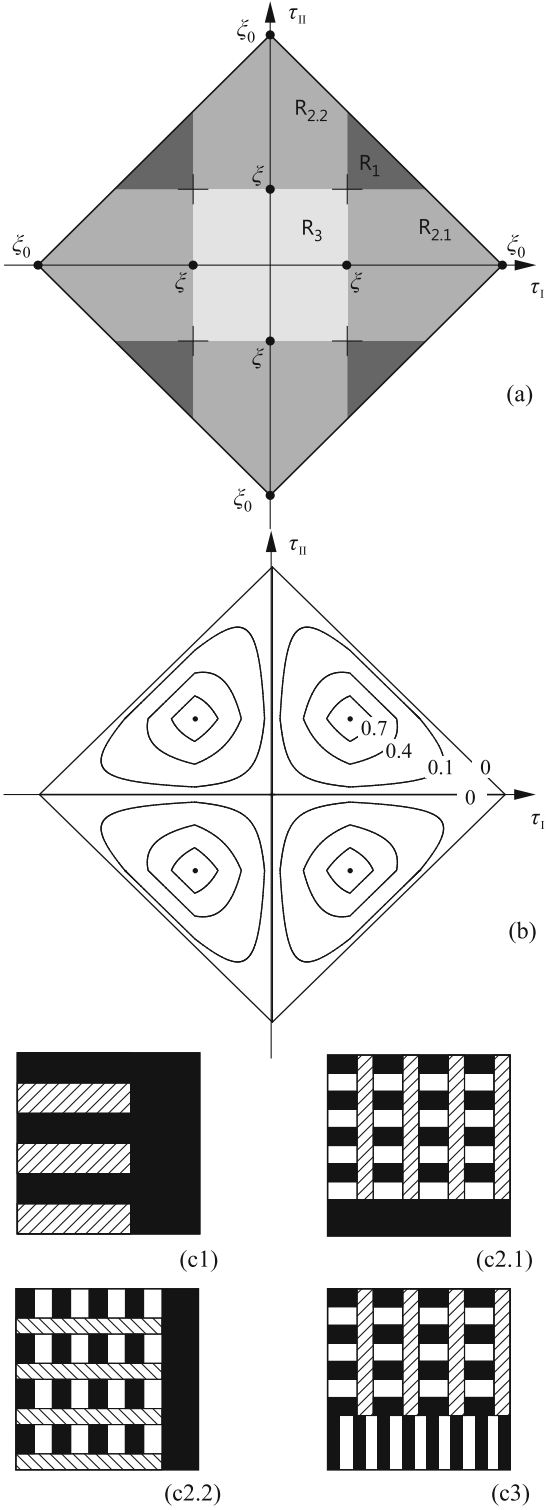}
\caption{{\bf a} Division of the composite region $\mathsf{R}(\alpha_A)$ into three subregions of optimality; {\bf b} Contour plot of $m_2^{\rm{max}}$,  $m_2^{\rm{max}}=1$ at the points $(\pm\xi, \pm\xi)$ where all subregions of optimality meet; {\bf c} Cartoons of optimal microstructures in respective subregions. All geometries represent laminates of a rank; the mixing of phases is hierarchical in scales. Black stripes denote phase 1, dashed -- phase 2 and white -- phase 3 (void). \label{fig1}}
\end{figure}

Weak material (phase 2) enters the optimal design when its cost is equal to the cost of optimal isotropic mixture of the strong material (phase 1) and void (phase 3). In this special case, the quotient $\alpha = \alpha_A$,
\be
\label{eq10}
\alpha_A = \dfrac{2\,C_1}{C_1+C_2}.
\ee
Indeed, one may check that if $\alpha=\alpha_A$ then the energy well $\Phi_2$ touches $QF_{13}$, i.e. 
\be
\label{eq10a}
2\,W_2(\gtau) + \alpha_A\gamma_1 = QF_{13}(\gtau),
\ee
for $\gtau = \gxi_i$, $i=1,\ldots, 4$, where
\be
\label{eq11a}
\begin{array}{ll}
\gxi_1 = (\xi, \xi), &\quad\gxi_2 = (-\xi, -\xi),\\[4pt]
\gxi_3 = (-\xi, \xi), &\quad\gxi_4 = (\xi, -\xi),
\end{array}
\ee
and
\be
\label{eq11b}
\xi =\dfrac{C_1}{C_1+C_2}\,\xi_0 .
\ee
Note that $\gxi_1 $, $\gxi_2 $ define pure spherical stress and $\gxi_3 $, $\gxi_4 $ are pure deviators. 

For $\alpha=\alpha_A$, the amount of material 2 is not uniquely defined because it can be arbitrarily interchanged with the mixture of phases 1 and 3. 
 Geometrically, the pa\-ra\-bo\-lo\-idal well $\Phi_2$ touches the quasiconvex envelope $QF_{13}$. If the cost of phase 2 is only slightly lower than the cost of the phase~1-phase~3 mixture, one wants to use the maximal amount of material 2 whenever it is possible. Therefore, optimal fraction of the weak material (phase 2) jumps from zero to some finite value for $\alpha=\alpha_A$.

From the above it follows that if $\alpha=\alpha_A$ then formula \eqref{eq08} remains valid but now $QF_{13}$ is additionally supported by $\Phi_2$. Hence, optimal macroscopic volume fractions cannot be uniquely determined for any $\gtau\in\mathsf{R}(\alpha_A)$, still defined by \eqref{eq08a}. The optimality of rank-2 laminates $L(13,1)$ characterized by \eqref{eq08b} is retained but one can as well use a fraction of material 2 to compile other optimal mixtures accumulating the same amount of stress energy for the same overall cost. Moreover, if the average stress $\gtau$ is represented by whichever tensor from \eqref{eq11a} then the cost of $L(13,1)$ is equal to the cost of pure phase 2 that is $\gamma_1\,m_1=\gamma_2$, which is equivalent to $m_1=\alpha_A$.

Composite region $\mathsf{R}(\alpha_A)$ splits into four subregions of optimality
\be
\label{eq11b1}
\begin{array}{ll}
\mathsf{R_1} &= \{\gtau : |\tau_{\rm{I}}|>\xi,\ |\tau_{\rm{II}}|>\xi, |\tau_{\rm{I}}|+|\tau_{\rm{II}}|<\xi_0\},\\[4pt]
\mathsf{R_{2.1}} &= \{\gtau : |\tau_{\rm{I}}|>\xi,\ |\tau_{\rm{II}}|<\xi, |\tau_{\rm{I}}|+|\tau_{\rm{II}}|<\xi_0\},\\[4pt]
\mathsf{R_{2.2}} &= \{\gtau : |\tau_{\rm{I}}|<\xi,\ |\tau_{\rm{II}}|>\xi, |\tau_{\rm{I}}|+|\tau_{\rm{II}}|<\xi_0\},\\[4pt]
\mathsf{R_3} &=  \{\gtau : |\tau_{\rm{I}}|<\xi,\ |\tau_{\rm{II}}|<\xi\} .
\end{array}
\ee 
In these regions, optimal macroscopic volume fraction of material 2 varies, $m_2\in (0, m_2^{\rm{max}})$, and
\be
\label{eq11b2}
m_2^{\rm{max}}=\begin{cases}
\dfrac{\xi_0-|\tau_{\rm{I}}| -|\tau_{\rm{II}}|}{\xi_0-2\,\xi} &\mbox{if}\ \ \gtau\in\mathsf{R_1},\\[8pt]
\dfrac{\xi_0-|\tau_{\rm{I}}| -|\tau_{\rm{II}}|}{\xi_0-\xi-|\tau_{\rm{II}}|}\,\dfrac{|\tau_{\rm{II}}|}{\xi} &\mbox{if}\ \ \gtau\in\mathsf{R_{2.1}},\\[8pt]
\dfrac{\xi_0-|\tau_{\rm{I}}| -|\tau_{\rm{II}}|}{\xi_0-|\tau_{\rm{I}}|-\xi}\,\dfrac{|\tau_{\rm{I}}|}{\xi} &\mbox{if}\ \ \gtau\in\mathsf{R_{2.2}},\\[8pt]
\dfrac{|\tau_{\rm{I}}|\,|\tau_{\rm{II}}|}{\xi^2}&\mbox{if}\ \ \gtau\in\mathsf{R_3}
\end{cases}
\ee
describes the maximal amount of the weak elastic phase that one can use to form the optimal high-rank, orthogonal laminate. The division of $\mathsf{R}(\alpha_A)$ and the contour plot of $m_2^{\rm{max}}$ are shown in Fig.\ \ref{fig1}(a) and Fig.\ \ref{fig1}(b) respectively. Optimal macroscopic fractions of phases 1 and 3 corresponding to \eqref{eq11b2} are given by
\be
\label{eq11b3}
\begin{array}{l}
m_1 = \dfrac{|\tau_{\rm{I}}| + |\tau_{\rm{II}}|}{\xi_0} - 2\,m_2^{\rm{max}}\,\dfrac{\xi}{\xi_0},\\[6pt]
m_3=1 - m_1 - m_2^{\rm{max}}.
\end{array}
\ee 
Examples of optimal microstructures are shown in Figs.\ \ref{fig1}(c1)-(c3). In these high-rank laminates, $\gtau^{(1)}$ and $\gtau^{(3)}$ (stress in phases 1 and 3) satisfy \eqref{eq09} and $\gtau^{(2)}$ (stress in phase 2) is given by one of the tensors in \eqref{eq11a}. Geometric parameters of optimal laminates are different in each subregion of $\mathsf{R}(\alpha_A)$. They are found by the technique used previously in \citep{Alb07, Che11} and \citep{Che13}. Roughly speaking, two types of equations are involved in the calculations: (i) formulae for average stress in two neighboring phases, and (ii) continuity of normal stress component on the interface between these phases.

\subsection{Low quotient of material costs}
\label{sec:low2}
\begin{figure}
\centering
\includegraphics[scale=1]{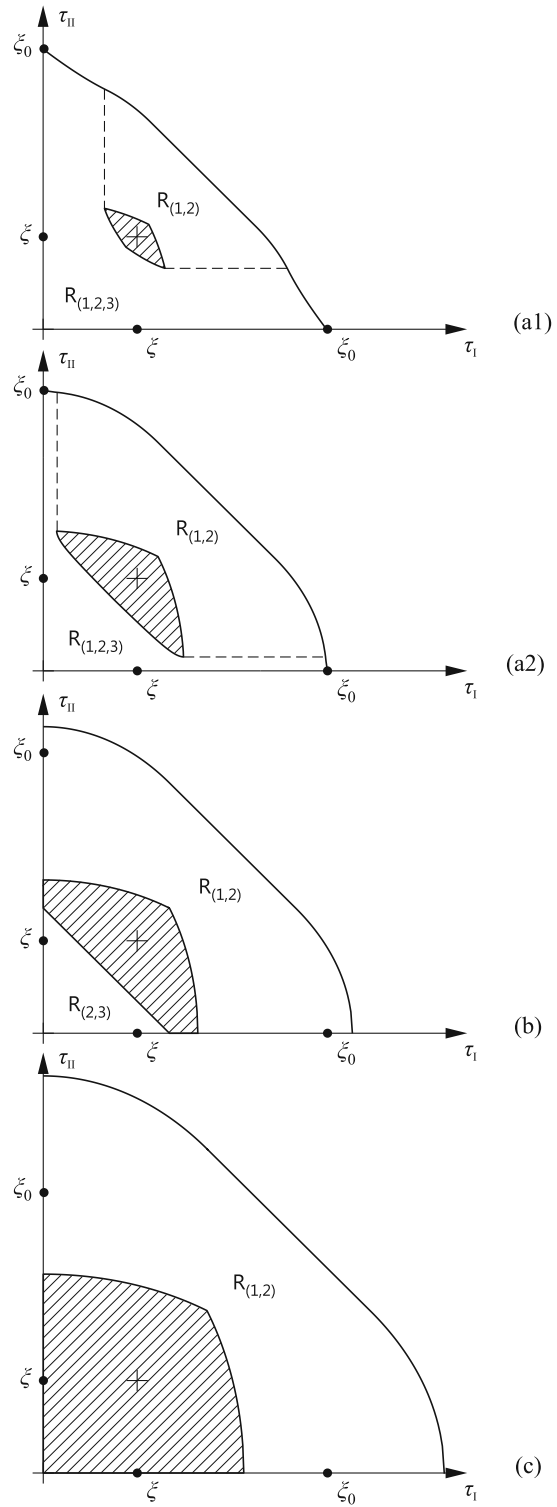}
\caption{Boundaries of the composite region $\mathsf{R}(\alpha)$ (top right quarter) for decreasing $\alpha$; {\bf a1} and {\bf a2} $\alpha_B<\alpha<\alpha_A$; {\bf b} $\alpha<\alpha_B$; {\bf c} $\alpha=0$. In each figure, the dashed area represents the domain where pure material 2 is optimal. The composite region is divided into subregions in which the optimality of high-rank laminates is conjectured, (see Sec.\ \ref{sec:low2} for the explanation of symbols $\mathsf{R_{(1,2)}}$, $\mathsf{R_{(2,3)}}$, and $\mathsf{R_{(1,2,3)}}$).\label{fig2}}
\end{figure}

Let us briefly outline the change in the topology of $\mathsf{R}(\alpha)$ when $\alpha<\alpha_A$, and the well $\Phi_2$ penetrates through the $QF_{13}$. The exact formulae for the quasiconvex envelope and details regarding geometry of optimal high-rank laminates are not reported here as they are subject to intensive ongoing research. Instead, we announce the qualitative results.

For $\alpha_B<\alpha<\alpha_A$, $\alpha_B=C_1/C_2$, the isolated zones appear around the points \eqref{eq11a}. Pure phase 2 is optimal in these zones and they expand as $\alpha$ decreases, see Figs.\ \ref{fig2}(a1)-(a2). The composite region $\mathsf{R}(\alpha)$ is divided into subregions with different optimal microstructures. We conjecture that rank-1 or rank-2 laminates of phases 1 and 2 are optimal in $\mathsf{R_{(1,2)}}$ and hierarchical mixtures of higher rank made of all three phases are optimal in $\mathsf{R_{(1,2,3)}}$. This hypothesis is based on the results presented in \citep{Che13}.

When the quotient of material costs further lowers, $\alpha\le\alpha_B$, the zones of optimality of pure material 2 merge and the composite region splits into two disconnected subdomains, see Fig. 2(b). In these subdomains, optimal composites are rank-2 laminates: $L(12,1)$ in subregion $\mathsf{R_{(1,2)}}$ and $L(23,2)$ in subregion $\mathsf{R_{(2,3)}}$. Three-material composites are not optimal for $\alpha\le\alpha_B$. 

Finally, when $\alpha= 0$, phase 3 (void) disappears from optimal design. In this case, rank-2 laminates $L(12,1)$ are optimal in the composite region, see Fig.\ \ref{fig2}(c). They can degenerate into rank-1 laminates for certain values of $\tau_{\rm{I}}, \tau_{\rm{II}}$.

\section{Results}
\label{sec:res}
\begin{figure*}
\centering
\includegraphics[scale=1]{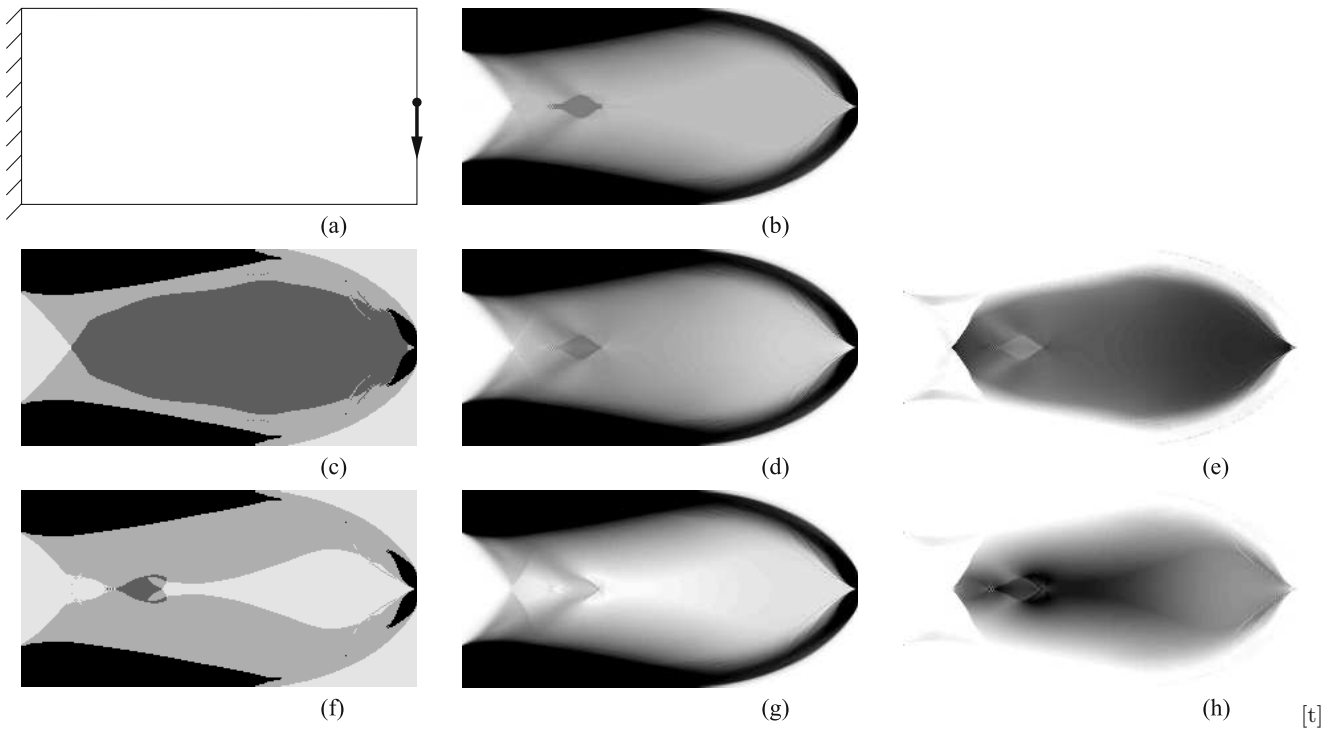}[t]
\caption{Optimal design of a cantilever; {\bf a} Design domain and loading; {\bf b} (two-phase design) Optimal distribution of material 1; {\bf c} -- {\bf e} (three-phase design, high relative cost $\alpha=\alpha_A$) Optimal microstructure regions; updated material 1 distribution; material 2 distribution;  {\bf f} -- {\bf h} (three-phase design, low relative cost $\alpha=\alpha_A$) Optimal microstructure regions; updated material 1 distribution; material 2 distribution. Black color in {\bf c} and {\bf f} denotes pure material 1 zone, others correspond to Fig.\ \ref{fig1}(a)\label{fig3}}
\end{figure*}
\begin{figure*}
\centering
\includegraphics[scale=1]{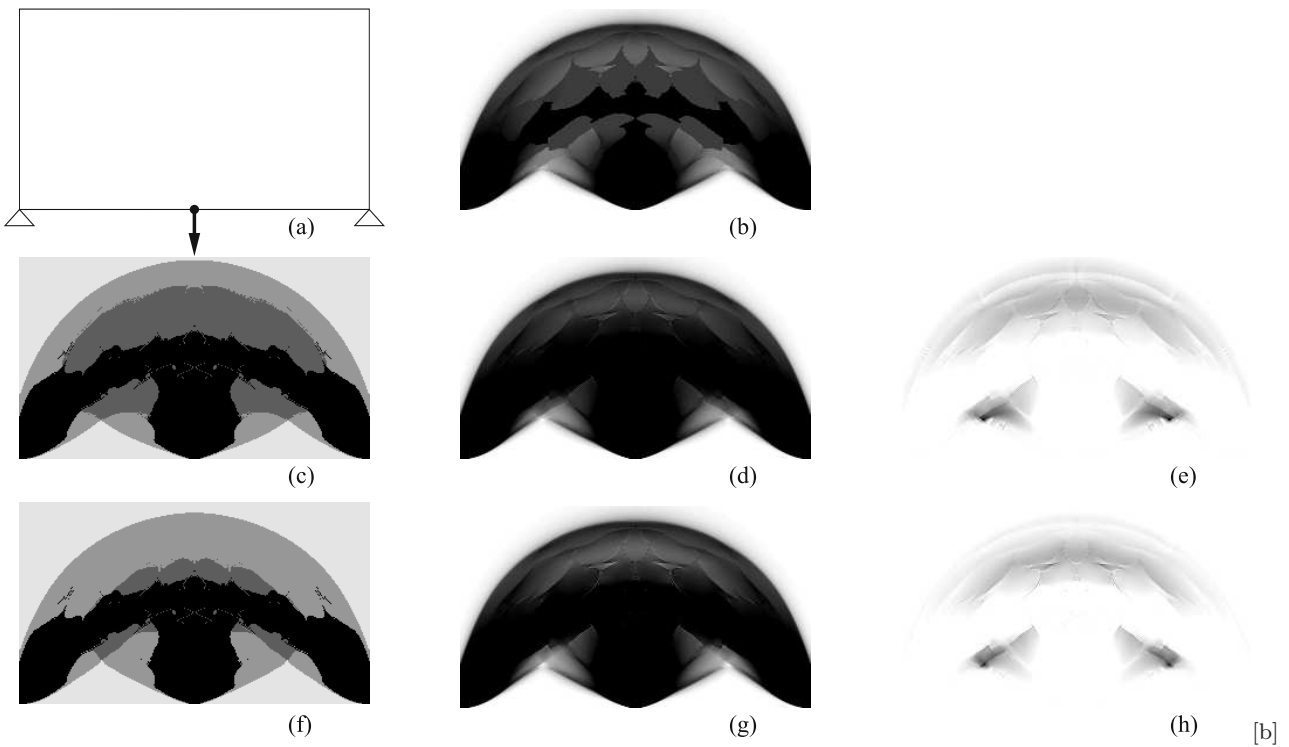}[b]
\caption{Optimal design of a bridge; {\bf a} Design domain and loading; {\bf b} (two-phase design) Optimal distribution of material 1; {\bf c} -- {\bf e} (three-phase design, high relative cost $\alpha=\alpha_A$) Optimal microstructure regions; updated material 1 distribution; material 2 distribution;  {\bf f} -- {\bf h} (three-phase design, low relative cost $\alpha=\alpha_A$) Optimal microstructure regions; updated material 1 distribution; material 2 distribution. Black color in {\bf c} and {\bf f} denotes pure material 1 zone, others correspond to Fig.\ \ref{fig1}(a) \label{fig4}}
\end{figure*}
Using the results in Sec.\ \ref{sec:critical2} we computed two standard examples of optimal design: a cantilever and a bridge. First, the topology optimization problem \eqref{eq06b} was solved for a given design domain $\Omega$ using the code by \citet{Dzi12}. In this way, the distribution of material 1 in $\Omega$ was found. Next, we determined the eigenvalues of the stress tensor at each $x\in\Omega$. Using this we computed the maximal amount of material 2 that can be used to replace a microstructure equivalent to the optimal $L(13,1)$. 

Optimal microstructures according to regions in Fig.\ \ref{fig1}(a) and the dependence of the results on the quotient $\alpha$ are denoted in Figs.\ \ref{fig3}, \ref{fig4}.

\section{Comments and Conclusions}
\label{sec:concl}
\begin{enumerate}
\item The strong material 1 tends to be placed close to the supports and the loading, while the weak phase 2 tends to concentrate in the regions where the stress tensor is closer to isotropic.  At the free boundary, the normal stress is zero and therefore phase 2 is not present. 

\item Phases 1 and 2 tend to be mixed. As in the material-void design, the regions of pure material 1 alternate with the composite zones forming ``ribs''. This provides the structural anisotropy and additional stiffness in the direction of maximal stress. 

\item We observe the ``almost void'' regions in the designs of the bridge and cantilever. However, the optimality conditions do not explain the sharp increase of the stress and the stiffness at some curves. We do not have a satisfactory explanation of this phenomenon. 

\item Comparing two- and three-phase designs, we observe a larger variety of the microstructures in the latter. For example, a second interior arc from material 2 is formed in the optimal bridge. We expect that this variety will be even more visible in the general situation, for lower quotient of material costs, $\alpha < \alpha_A$. 

\item The three-phase optimal layout is not unique when $\alpha=\alpha_A$. For certain values of the average stress, one has the option to replace a $L(13,1)$ microstructure with pure material 2, or to include material 2 into the composition. This is often preferable because the structures where material 2 replaces some part of void are more stable and therefore have a better response to the variations in loading. 
\end{enumerate} 

\begin{acknowledgement}
Part of this work was supported thro\-ugh the Research Grant no 2013/11/B/ST8/04436 financed by the National Science Centre (Poland), entitled: \emph{Topology optimization of engineering structures. An approach synthesizing the methods of: free material design, composite design and Michell-like trusses} (Grzegorz Dzier\.zanowski).
\end{acknowledgement}

\end{document}